\documentclass[namedreferences]{SolarPhysics}
\usepackage[optionalrh]{spr-sola-addons} 
\usepackage{graphicx}        
\usepackage{url}             

\newcommand{\eg}{{\it e.g.}}
\newcommand{\ie}{{\it i.e.}}

\newcommand{\adv}{    {\it Adv. Space Res.}}

\newcommand{\aap}{    {\it Astron. Astrophys.}}

\newcommand{\apj}{    {\it Astrophys. J.}}
\newcommand{\apjl}{   {\it Astrophys. J.}}

\newcommand{\araa}{   {\it Ann. Rev. Astron. Astrophys.}}

\newcommand{\jgr}{    {\it J. Geophys. Res.}}

\newcommand{\pasj}{   {\it Pub. Astron. Soc. Japan}}

\newcommand{\solphys}{{\it Solar Phys.}}

\begin{document}
\begin{article}
\begin{opening}

\title{Reconnection of a kinking flux rope triggering the ejection of a
       microwave and hard X-ray source}
\subtitle{I. Observations and Interpretation}

\author{M. \surname{Karlick\'y}$^{1}$,
        B. \surname{Kliem}$^{2,3,4}$}
        \runningauthor{Karlick\'y and Kliem}
        \runningtitle{Reconnection in kinking flux rope ejection I}

\institute{$^{1}$ Astronomical Institute, Academy of Sciences of
                  the Czech Republic, 251 65 Ond\v{r}ejov, Czech Republic
                  \email{karlicky@asu.cas.cz}\\
           $^{2}$ Universit\"{a}t Potsdam, Institut f\"{u}r Physik und
                  Astronomie, 14476 Potsdam, Germany\\
           $^{3}$ University College London, Mullard Space Science Laboratory,
                  Holmbury St.~Mary, Dorking, Surrey, RH5 6NT, UK\\
           $^{4}$ Naval Research Laboratory, Space Science Division,
                  Washington, DC 20375, USA}

\date{Received 30 July 2009; accepted 5 July 2010}

\begin{abstract}
Imaging microwave observations of an eruptive, partially occulted solar flare
on 18 April 2001 suggest that the global structure of the event can be
described by the helical kink instability of a twisted magnetic flux rope. This
model is suggested by the inverse gamma shape of the source exhibiting crossing
legs of a rising flux loop and by evidence that the legs interact at or near
the crossing point. The interaction is reflected by the location of peak
brightness near the crossing point and by the formation of superimposed compact
nonthermal sources most likely at or near the crossing point. These sources
propagate upward along both legs, merge into a single, bright source at the top
of the structure, and continue to rise at a velocity $>\!1000$~km\,s$^{-1}$.
The compact sources trap accelerated electrons which radiate in the radio and
hard X-ray ranges. This suggests that they are plasmoids, although their
internal structure is not revealed by the data. They exhibit variations of the
radio brightness temperature at a characteristic time scale of $\sim\!40$~s,
anti-correlated to their area, which also support their interpretation as
plasmoids. Their propagation path differs from the standard scenario of
plasmoid formation and propagation in the flare current sheet, suggesting the
helical current sheet formed by the instability instead.

\end{abstract}

\keywords{Sun: flares -- Sun: radio radiation}

\end{opening}

\section{Introduction}
\label{s:introduction}

Many solar eruptions suggest that a kinking magnetic flux rope is
involved. The characteristic helical shape acquired by a rising and kinking
(writhing) flux loop manifests itself as a rotation of the top part of
the loop leading either to a crossing of the legs (also occasionally referred
to as an inverse gamma shape) or to an O shape of the loop, depending on the
perspective of the observer (\eg, \opencite{Ji&al2003};
\opencite{Romano&al2003}; \opencite{Gilbert&al2007}; \opencite{Cho&al2009}). A
role for the helical kink instability in initiating eruptions
(\opencite{Torok&Kliem2005}; \opencite{Fan2005}) is nevertheless debated, since
many observations and recent modeling of filaments prior to eruption indicate
subcritical values of the twist (\eg, \opencite{Bobra&al2008}). It is possible,
however, that much of the twist in erupting flux ropes is built up in the early
phases of the eruption through the addition of current-carrying flux by
reconnection under the rope (\eg, \opencite{Lin&al2004}). Moreover, filaments
outline only a small fraction of the current-carrying flux in an active region,
so that part of the existing twist may remain undetected. An example is the
filament in NOAA active region (AR) 10930, which did not appear
particularly twisted \cite{Williams&al2009}, although the sunspot near one end
of the filament did rotate by $540^\circ$ before the region erupted
\cite{Min&Chae2009}. A second reason for questioning the occurrence of the
instability lies in the fact that the writhing and apex rotation of the flux
loop may also be caused by the shear field component of the ambient field
(pointing along the polarity inversion line of the photospheric flux) if a
process other than this instability accelerates the loop upwards
\cite{Isenberg&Forbes2007}. The upward acceleration may be caused by the torus
instability \cite{Kliem&Torok2006, Torok&Kliem2007}. On the other hand, many
erupting filaments exhibit clear indications for \emph{both}, an overall
helical shape, \ie, a writhed axis, and a winding of the threads in the
filament about the axis, \ie, twist; this combination supports the occurrence
of the helical kink mode strongly. A particularly clear case was described by
\inlinecite{Romano&al2003}, who estimated a twist of $\sim\!10\pi$ in a
filament early in its eruption. The twist decreased subsequently, as expected
for a helical kink. Indications for high twist values in prominences near the
onset of an eruption, including estimates made before the onset and in
agreement with stability estimates for the helical kink mode, were presented by
\inlinecite{Vrsnak&al1991} and \inlinecite{Vrsnak&al1993}.

Usually, the leg crossing is a pure projection effect, expected for any
writhing loop independent of the cause of the writhing, provided the observer
has an appropriate perspective on the eruption. For moderately supercritical
values of the twist, $\Phi\sim(3\mbox{--}5)\pi$, such that the loop develops
about one helical turn in total (\ie, the perturbation has a normalized axial
wavenumber $k'=kl/2\pi\sim1$, where $l$ is the length of the flux loop), the
helical kink actually drives the legs \emph{away} from each other. The front
leg is then moving toward the observer and the rear leg is moving away (see,
\eg, Fig.~4 in \opencite{Torok&Kliem2005} and Fig.~14 in
\opencite{Green&al2007}). A completely similar motion of the legs results from
the writhing that is caused by the shear field. However, the eruptive flare on
18 April 2001 in AR~9415, whose microwave emissions are studied in the present
paper, did not only develop an inverse gamma shape but also gave indications
that the crossing legs indeed interacted, producing superimposed sources and
nonthermal particles. The superimposed, relatively compact microwave and hard
X-ray sources originated near the crossing point and propagated along the legs
to the top of the erupting loop. We point out, and demonstrate through
numerical simulation in the companion paper (\opencite{Kliem&al2010}; hereafter
Paper~II), that the interaction of crossing legs can result from the helical
kink instability of a flux rope if modes with $k'\sim2$ are dominant as a
result of high initial twist in the range $\Phi\gtrsim6\pi$.

The superimposed, propagating sources of the microwave and hard X-ray emission
represent a second interesting aspect of the event under investigation. These
``blobs'' trap the accelerated particles, which is a known property of
plasmoids (\eg, \opencite{Karlicky&Barta2007}). X-ray sources of the same or
similar nature have been known for a long time. They have first been recognized
in association with so-called long-duration events (\eg, Hudson, 1994;
\v{S}vestka {\it et al.}, 1995). Similar X-ray plasma ejections have later been
found also in impulsive, compact flares (\eg, Shibata {\it et al.}, 1995;
Nitta, 1996; Tsuneta, 1997; Ohyama and Shibata, 1997; Ohyama {\it et al.},
1997). Based on this fact, Ohyama and Shibata (1998) and Shibata and Tanuma
(2001) proposed a general scenario of plasmoid formation and ejection. They
assumed that an extended (flare) current sheet is formed below a rising
magnetic flux rope. Magnetic islands (plasmoids) are formed in this current
sheet, due to the tearing-mode instability, and move along the sheet, possibly
coalescing into larger units. Furthermore, a close temporal association between
plasmoid ejections and so-called drifting pulsation structures was found. These
are observed in dynamic radio spectra in the $\sim600\mbox{--}2000$~MHz range.
Their origin was related to quasi-periodic episodes of particle acceleration in
an intermittent (``bursty'') mode of reconnection in the flare current sheet,
combined with the trapping of the accelerated particles in the resulting
plasmoids (Kliem, Karlick\'y, and Benz, 2000; Karlick\'y, 2004; Karlick\'y,
F\'arn\'{\i}k, and Krucker, 2004; Karlick\'y and B\'arta, 2007). In all these
cases, the formation and motion of plasmoids is supposed to occur in the
vertical flare current sheet \emph{below} a rising flux rope (which often
contains a filament or prominence). The superimposed sources in the 18 April
2001 flare, however, did clearly propagate along the legs to the top of the
rising loop.

In this paper we present a detailed analysis of the microwave
observations in the course of the eruptive flare. The imaging data were
taken by the Nobeyama Radioheliograph at 17 and 34~GHz. They are
complemented by light curves from the Nobeyama Radiopolarimeters and by
a dynamic spectrum of the dm--Dm emission obtained by the HiRAS
spectrograph. This event was already studied by
\inlinecite{Hudson&al2001}, who investigated the fast coronal ejection
in hard X-rays, observed in the 23--53 keV energy range by the
\textsl{Yohkoh} Hard X-ray Telescope, joint with the strongest moving
source mapped by the Nobeyama Radioheliograph. However, they did not
analyze the evolution that led to the formation of the ejected source,
which is suggestive of a plasmoid or similar structure in, or in the
vicinity, of an erupting, kinking flux rope, different from the
classical picture of formation in the flare current sheet.

Paper~II presents MHD simulations of the helical kink instability of a
highly twisted flux rope, which corroborate the model of reconnection
between flux rope legs suggested by the present data analysis.

\section{Observations and Interpretation}
\label{s:observations}

\begin{figure}[t]\begin{center}
\includegraphics[scale=0.8]{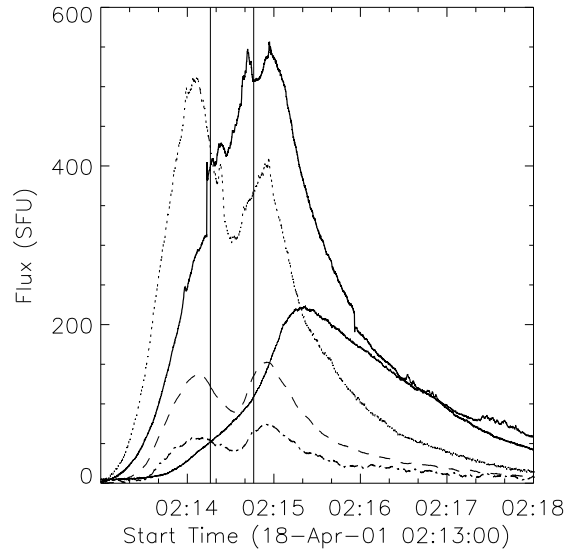}
\caption{Centimeter--decimeter time profiles of the 18 April 2001 flare
observed by the Nobeyama Radiopolarimeters at the frequencies 1~GHz (thin
solid line), 2~GHz (thick solid), 3.75~GHz (dotted), 9.4~GHz (dashed),
and 17~GHz (dot-dashed). Background fluxes are subtracted.
The two vertical lines mark the times of the first and second (main)
23--33~keV hard X-ray flux maxima given in Hudson {\it et al.} (2001).}
\label{f:lightcurves}
\end{center}\end{figure}

\begin{figure}[t]\centering
\includegraphics[scale=0.58]{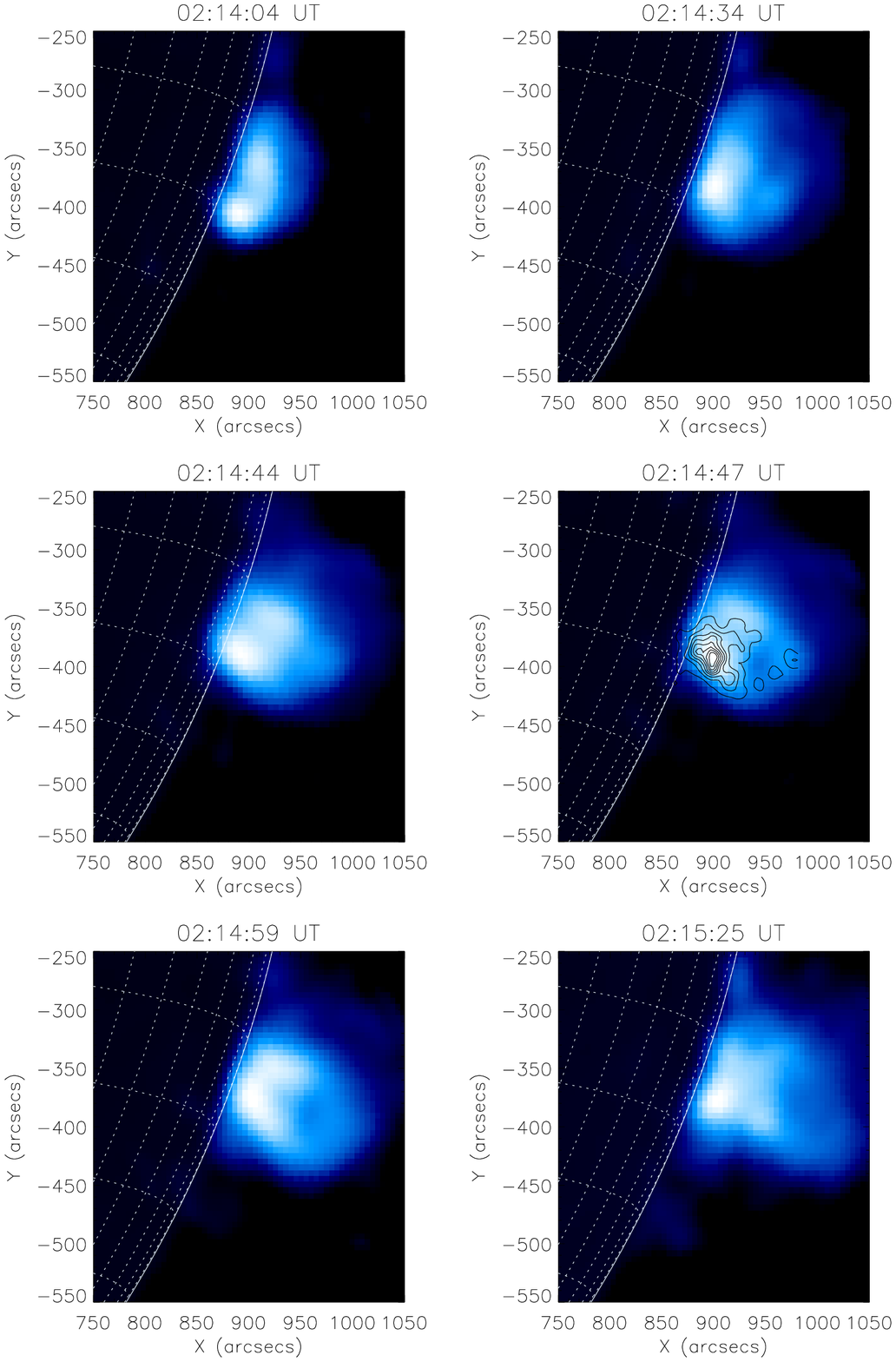}
\caption{17~GHz Nobeyama Radioheliograph images of the 18 April 2001
flare comprising the period of the two peaks in the microwave and hard
X-ray time profiles. Logarithmic brightness temperature in the range
$8000\mbox{--}10^6$~K is displayed, and 23--33~keV contours of the hard
X-ray source at the time of peak brightness are overlaid on the
simultaneous 17~GHz image. The sequence shows the rise of the main,
inverse gamma shaped microwave source above the solar limb and the
proximity of the brightest microwave and hard X-ray emission to the
crossing point.}
\label{f:leg_crossing}
\end{figure}

\begin{figure}[t]                                                
\centering
\includegraphics[width=.7\textwidth]{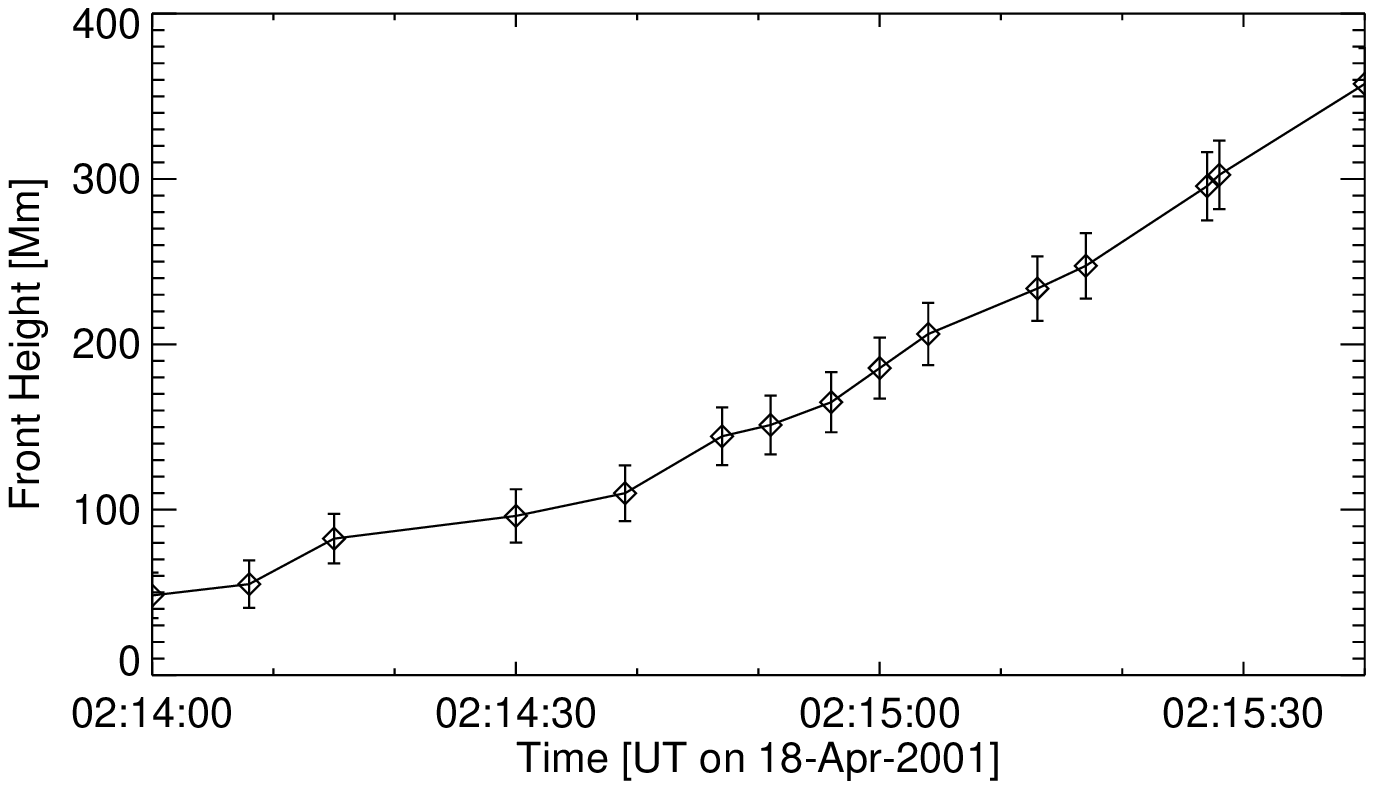}  
\caption{
 Rise profile of the front edge of the inverse gamma shaped main
 microwave source at 17~GHz.}
\label{f:accel}
\end{figure}

The 18 April 2001 flare, classified as \textsl{GOES}~C2.2, occurred in NOAA
AR~9415 at a location $\approx\!26^\circ$ behind the west limb, corresponding
to a radial occultation height of $\approx\!110$~arcsec \cite{Hudson&al2001}.
Its temporal evolution observed by the Nobeyama Radiopolarimeters at 1, 2,
3.75, 9.4, and 17~GHz is shown in Figure~\ref{f:lightcurves}. The higher
frequencies reveal two peaks, around 02:14:10 and 02:14:55~UT, whereas only the
second, main peak is seen at the lower frequencies (which show a steeper rise
at the time of the first peak). This structure is due to the rise of different
parts of the microwave source above the limb as well as to intrinsic variations
in the source, and it is similar to the light curves in the hard and soft X-ray
ranges, respectively (see Fig.~1 in \opencite{Hudson&al2001}). The main peak
shows a systematic delay toward lower frequencies, which can be related to a
change from free-free emission at microwaves to plasma emission at decimetric
wavelengths (see the description of the dynamic radio spectrum below).

We searched for any phenomena at earlier times, which might yield clues
to the onset of this behind-limb event. The \textsl{SOHO}/EIT images at
01:25:58, 01:35:19, 01:48:12, 02:00:11, and 02:12:10~UT show no
structural changes above the solar limb at position angles near the
eruption. Only a weak jet was found about 100~arcsecs northwards from
this location in the 01:48:12~UT image, too early to provide an
immediate trigger of the fast eruption.

Similarly, the EIT images taken during and after the event at 12~min
cadence in the 195~{\AA} band do not reveal any flare-related features,
most likely because of the behind-limb location.

\begin{figure}[t]\begin{center}
\includegraphics[scale=0.6]{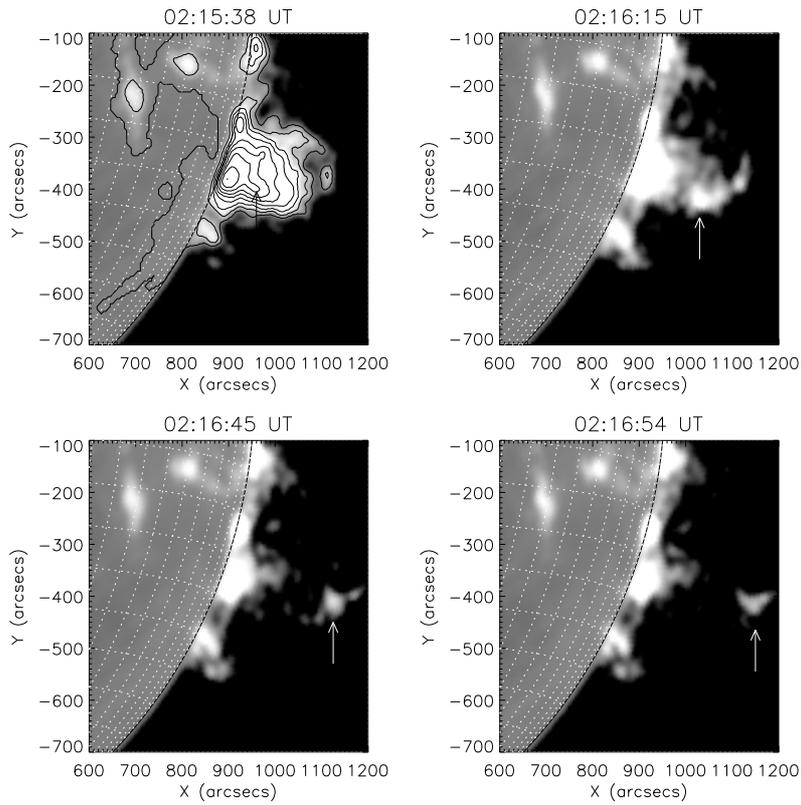}
\caption{17~GHz Nobeyama Radioheliograph images comprising the period of
the rapidly rising, superimposed compact sources. Logarithmic brightness
temperature in the range 6000--20,000~K is displayed in grayscale, with
contours of logarithmic brightness in the range $8000\mbox{--}10^6$~K
overplotted on the first image. White arrows in panels~2--4 mark the brightest
compact source. The black arrow in panel~1 marks the position of the brightest
23--33~keV hard X-ray source at the same time, which coincides with the
brightest compact microwave source (see Figs.~2 and 3 in Hudson {\it et al.},
2001).}
\label{f:main_plasmoid}
\end{center}\end{figure}

\begin{figure}[t]\begin{center}
\includegraphics[scale=0.6]{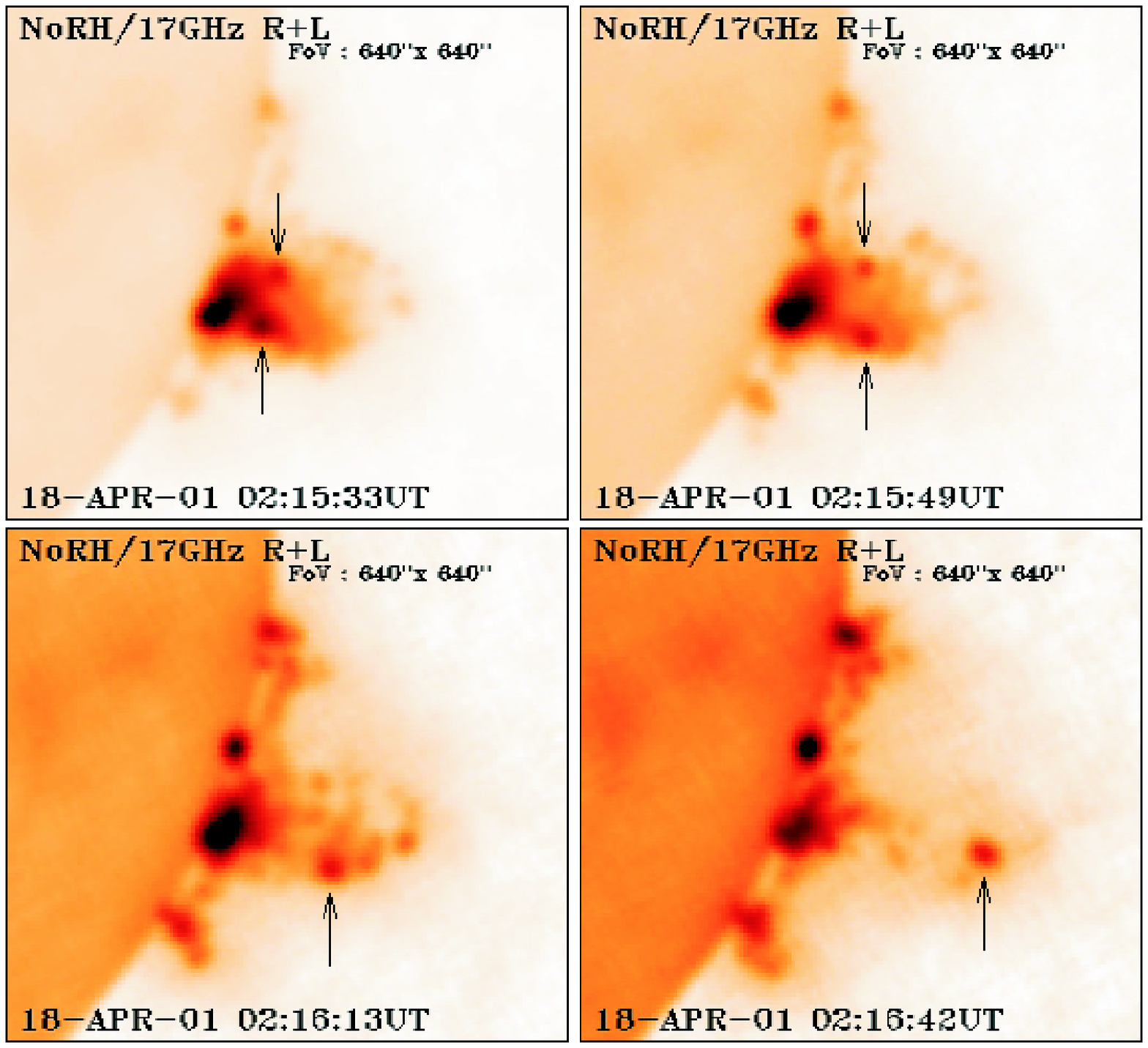}
\caption{Similar to Figure~\ref{f:main_plasmoid}, with different
weighting of the visibilities to emphasize small scale structure
(courtesy of M.~Shimojo). A series of upward moving compact sources,
arranged along the upper loop of the main, inverse gamma shaped source,
is now apparent, the brightest of which are marked by arrows. All
compact sources merge eventually at the top of the main microwave
source (final panel), continuing the rapid ascent.}
\label{f:plasmoid_merging}
\end{center}\end{figure}

The evolution of the flare observed by the Nobeyama Radioheliograph at 17~GHz
with a spatial resolution of 10~arcsec in the interval 02:14:04--02:15:25~UT is
presented in Figure~\ref{f:leg_crossing}. Overall, the microwave source
exhibits a rapid evolution, indicative of a rather unstable configuration. By
about 02:14:04~UT, the time of the first microwave and hard X-ray peak, the
source has developed into the top part of a single flare loop. Then it changes
to concave shape, accompanied by weak, diffuse emission at greater heights
(02:14:34~UT). With the second rise of the flux, the diffuse emission turns
into a clearer loop shape, now in the form of a complete ellipse (02:14:59~UT).
Next a cross appears above the limb as the ellipse rises further (02:15:25~UT).
This configuration is similar to the inverse gamma shape observed in strongly
kinking filaments. The peak microwave emission is located very near the
crossing point, and the rise of the crossing point above the limb is associated
with the main peak of the 17~GHz light curve (Figure~\ref{f:lightcurves}).
After $\sim$\, 02:16~UT the upper part of the inverse gamma shaped main
microwave source faded to fall out of the dynamic range of the radio images,
but the source remained bright at the limb near the original position of the crossing
point, as if the crossing point did not participate in the subsequent expansion.
Rather, the location of peak brightness at the limb is consistent with
a slight retraction of the crossing point to or even behind the limb
(Figures~\ref{f:main_plasmoid} and \ref{f:plasmoid_merging}).

For comparison, the contours of the 23--33~keV source at the time of the main
hard X-ray flux peak (02:14:47~UT) are overlaid on the simultaneous microwave
image in Figure~\ref{f:leg_crossing}. The overlay shows that the hard X-ray
source emerges from about the position of the crossing legs of the rising loop.
At this time it is not tied to the loop top. This renders an association with a
shock, which might be driven by the rising loop, to be unlikely. Also, the
motion of the X-ray source changes from a slow rise, still staying near the
limb (consistent with occultation of the source's main part), to a fast one at
about this time. The fast rise can be fit by a linear function to yield a
velocity of $\approx\!930$~km\,s$^{-1}$; see Fig.~3 in
\inlinecite{Hudson&al2001}. In other words, the main acceleration of this
source ended shortly after the main hard X-ray peak, as is typical of fast
ejections (\eg, \opencite{Zhang&Dere2006}; \opencite{Temmer&al2008}).
\inlinecite{Hudson&al2001} also show that the relatively compact hard X-ray
source is nonthermal, with a power-law spectrum in the range
$\sim\!23\mbox{--}93$~keV, throughout the period it could be detected
($\approx$\,02:13:50--02:17~UT).

In order to estimate the main acceleration of the inverse gamma shaped
microwave source, images sharper than those in Figure~\ref{f:leg_crossing} are
required. These are obtained by employing a different weighting of the
visibilities provided by the interferometer (see
Figure~\ref{f:plasmoid_merging} below). The front height of the source vs.\
time and the corresponding errors can then roughly be estimated in the period
02:14--02:15:40~UT (Figure~\ref{f:accel}). Although the data scatter
substantially, it is obvious that the acceleration extended up to
$\sim$\,02:15:20~UT and peaked at about 02:14:45~UT, very close to the time of
peak hard X-ray flux; even a secondary jump in velocity at the time of the
first hard X-ray maximum can be seen.

Isolated rising blobs of microwave emission, superimposed on the fading inverse
gamma shaped main source, became discernible soon after the crossing point had
appeared above the limb. We term these sources ``compact'', although they are
not necessarily below the resolution limit of the interferometer at all times.
Two of them occurred slightly above the cross, one in each of the crossed legs,
at about 02:15:33~UT (marked by arrows in Figures~\ref{f:main_plasmoid} and
\ref{f:plasmoid_merging}), which is shortly after the main acceleration of the
ejecta. The brighter one, in the southern leg of the erupting loop, was
co-spatial with the nonthermal hard X-ray source (see Figs.~1--3 in
\opencite{Hudson&al2001}). The images between 02:15:49 and 02:16:15~UT in
Figures~\ref{f:main_plasmoid} and \ref{f:plasmoid_merging} show that even a
series of compact sources formed, their location being consistent with an
arrangement along the rising loop above the crossing point. The sources moved
upward and merged near the top of the loop, where a single source remained
after 02:16:35~UT (see the images after 02:16:15~UT in
Figures~\ref{f:main_plasmoid} and \ref{f:plasmoid_merging}). The final compact
source continued its rise at high speed, evolving into a fast coronal mass
ejection (CME).

\begin{figure}[t]\begin{center}
\includegraphics[scale=0.6]{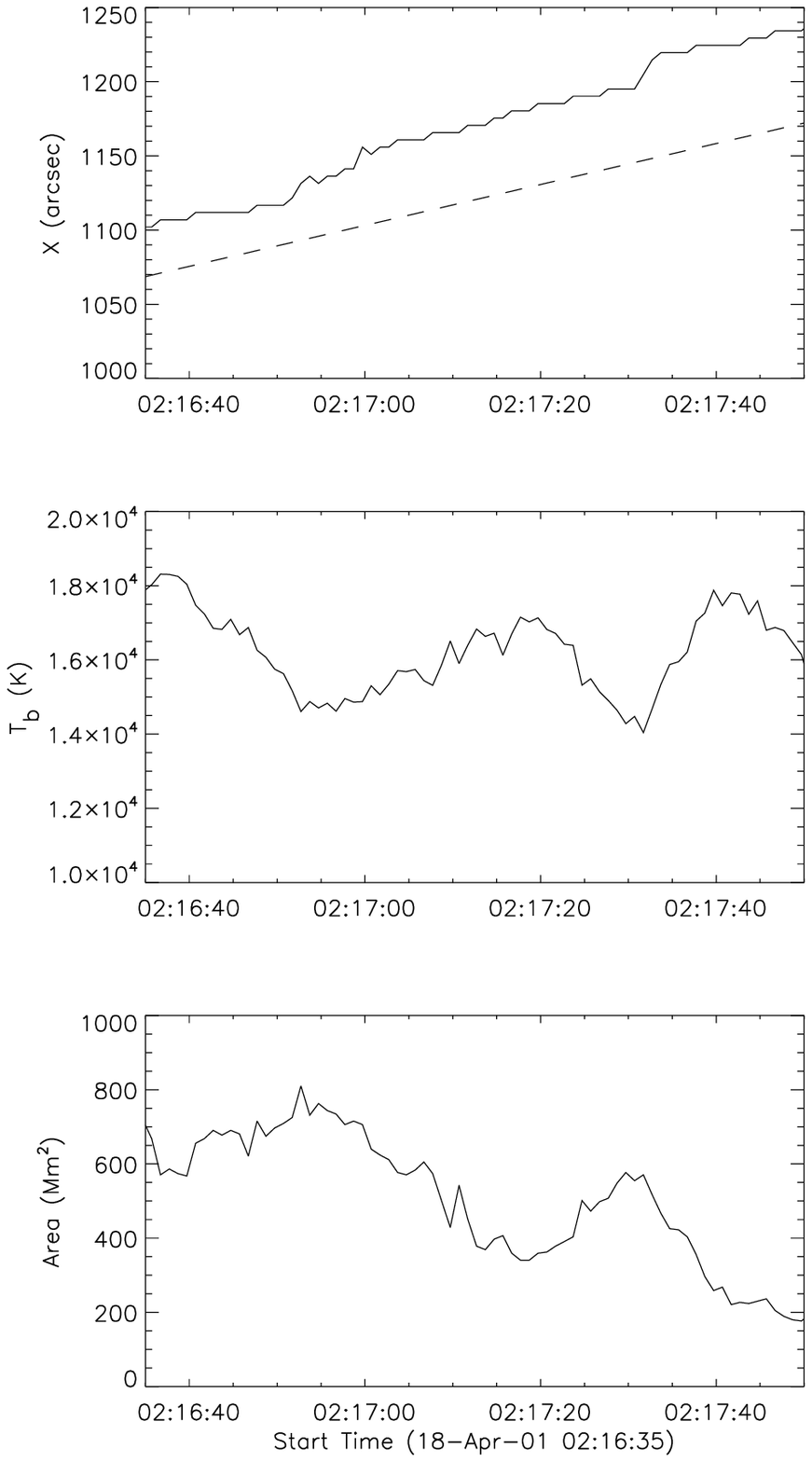}
\caption{Evolution of the final compact 17~GHz source in the interval
02:16:35--02:17:50~UT, \ie, after the merging of all rising blobs into a single
source. {\it Top:} Solid line: east-west source location. For comparison, the
dashed line represents a velocity of 1000~km\,s$^{-1}$. {\it Middle:} Maximum
brightness temperature. {\it Bottom:} Source area at half the peak brightness
temperature.}
\label{f:timeprofiles}
\end{center}\end{figure}

\begin{figure}[t]\begin{center}
\includegraphics[angle=-90,width=\textwidth]{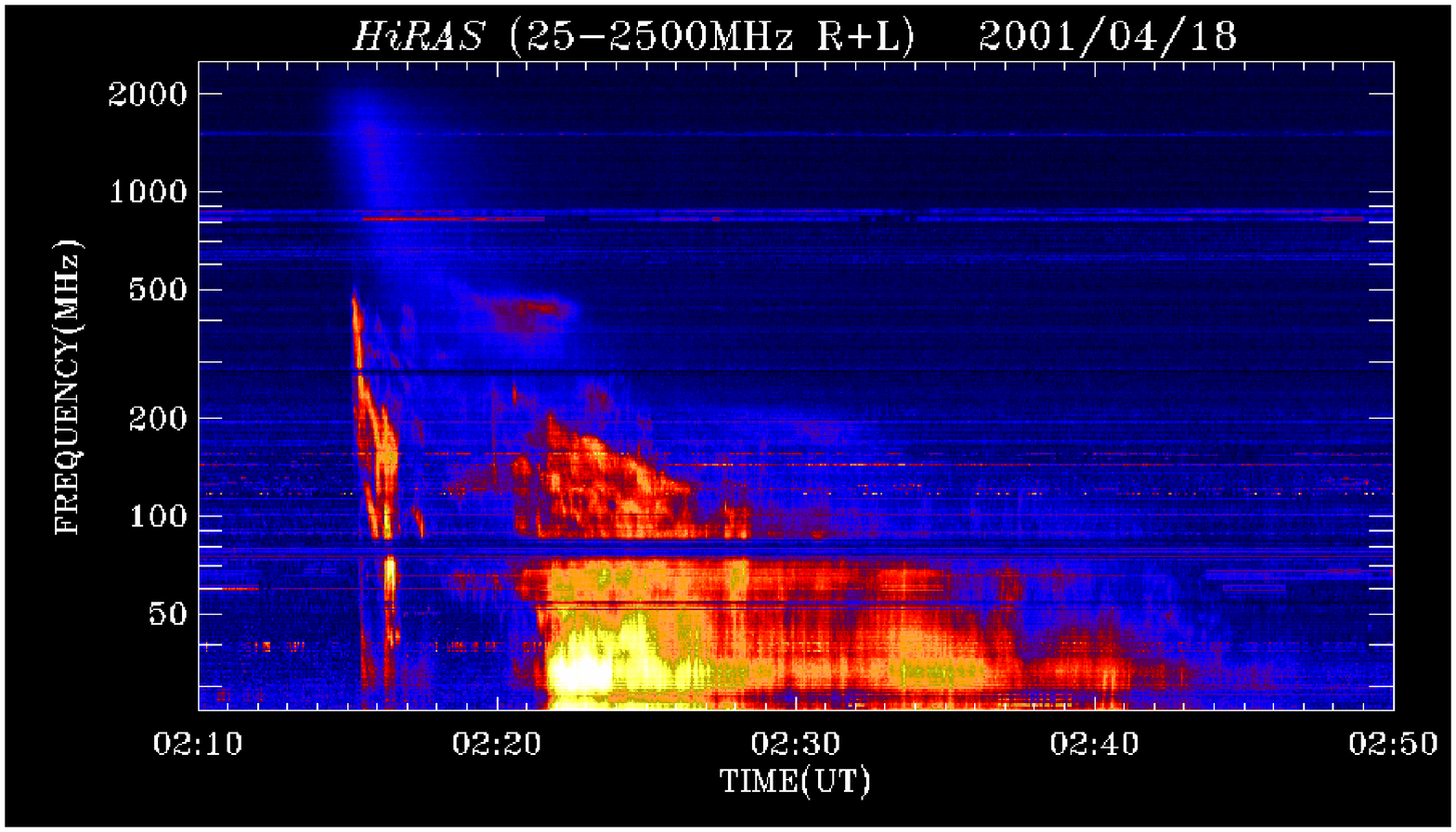}
\caption{Dynamic spectrum from the HiRAS Spectrometer, showing
the bright slow-drift (type II) burst indicative of a CME-driven coronal shock
wave (starting at $\approx$\,[02:20 UT, 300 MHz]), the weaker drifting
continuum likely associated with the ejected microwave and hard X-ray sources
(starting at $\approx$\,[02:14 UT, 2000 MHz]), and the group of fast-drift
(type III) bursts (starting at $\approx$\,[02:15 UT, 500 MHz]).}
\label{f:dynamic_spectrum}
\end{center}\end{figure}

Figure~\ref{f:timeprofiles} displays the east-west position of the final
compact source vs.\ time, along with its maximum brightness
temperature and its area at half the peak brightness temperature. It is seen
that the source continued its westward rise at the high projected speed of
$\gtrsim\!1000$~km\,s$^{-1}$ obtained from the hard X-ray images up to
02:15:45~UT in \inlinecite{Hudson&al2001}. The westward rise is close to the
initial direction of the CME and at about half the projected velocity of the
CME leading edge (estimated to be $\approx\!2500$~km\,s$^{-1}$ in the LASCO CME
catalogue at NASA's CDAW data center,
\texttt{http://cdaw.gsfc.nasa.gov}). This lies within the range of core
and leading-edge velocity ratios observed in fast CMEs \cite{Maricic&al2009}.
As expected from the high rise velocity, the formation of a large-scale coronal
shock is indicated by an associated slow-drift (type II) burst commencing at
$\approx$\,02:20~UT near 300~MHz; see the dynamic spectrum in
Figure~\ref{f:dynamic_spectrum}.

In addition to the bright type II burst, a weaker continuum can be discerned in
the dynamic spectrum from about 02:14~UT onward. The continuum had a starting
frequency near 2~GHz and drifted to lower frequencies at an intermediate rate,
such that its trace arrived near but above the frequency of the type II burst
(by a factor 1.5--2) when that burst commenced. Its sharp onset in the 2~GHz
time profile, as well as the delayed onset in the 1~GHz profile, are also
clearly visible in Figure~\ref{f:lightcurves}. It is plausible to assume that
the drifting continuum is caused by plasma emission (near the electron plasma
frequency or its harmonic) associated with the appearance of the leg crossing
above the limb and with the compact microwave/hard X-ray sources. This
interpretation implies high densities, starting slightly above
$10^{10}$~cm$^{-3}$ (adopting the harmonic assumption) and gradually decreasing
to $n_e\sim\!7$, 4, and $1\times10^9$~cm$^{-3}$ by 02:15:30, 02:16, and
02:17~UT, respectively. These values are consistent with the independent
estimate of the source density of $\lesssim\!4\times10^9$~cm$^{-3}$ at
$\approx$\,02:15:45~UT in \inlinecite{Hudson&al2001}, who used the expression
for the brightness temperature of free-free emission joint with estimates of
source temperature and depth of the brightest compact 17~GHz source (see below
for the expression). High densities support the assumption that the compact
sources are plasmoids.

\begin{figure}[t]                                                
\centering
\includegraphics[scale=0.4]{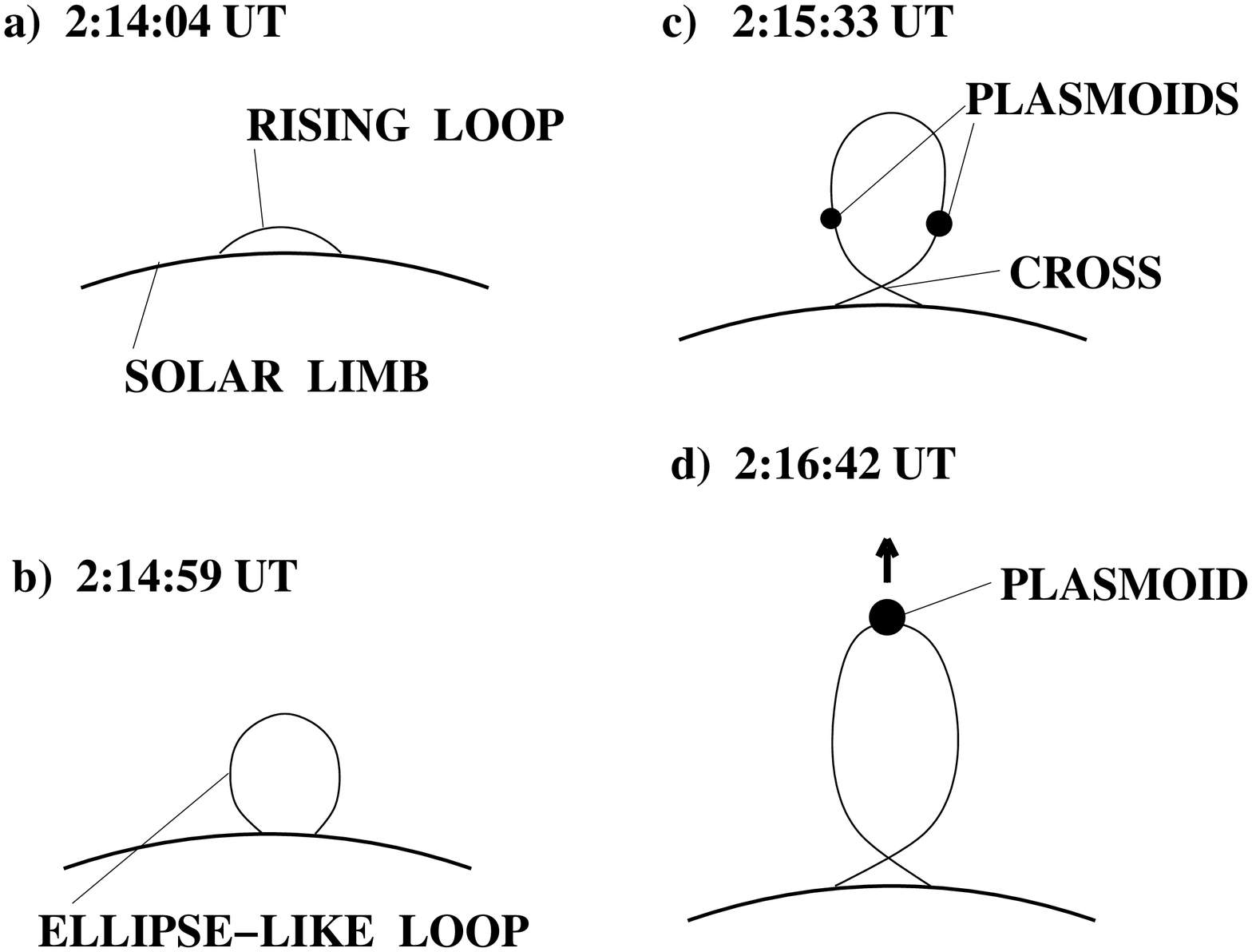}
\caption{Schematic summary of the flare evolution at 17~GHz. The compact
rising sources, the brightest of which is co-spatial to the hard X-ray
source, are suggested to be plasmoids.}
\label{f:schematic}
\end{figure}

\begin{figure}[t]                                                
\centering
\includegraphics[scale=0.78]{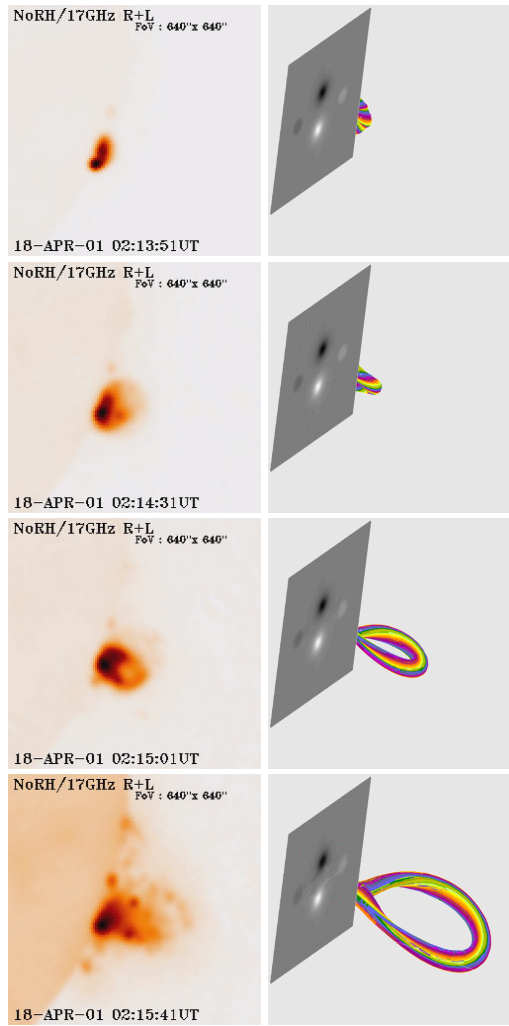} 
\caption{Comparison of the evolving 17~GHz source with the shape of a
simulated kink-unstable flux rope displayed at the same viewing angle (tilted
away from the observer by 26$^\circ$, so that the magnetogram,
$B_z(x,y,0,t)$, is seen from the bottom side). The flux rope has an initial
twist of $6\pi$ and is also torus unstable. Snapshots of field lines in
the core of the flux rope, started at circles
at the apex, are shown at the times $t=9.1$, 18.0, 23.6, and $28.6\tau_A$
(Alfv\'en times), corresponding to apex heights $h_\mathrm{a}=1.1$, 1.4, 2.8,
and $4.5h_0$, respectively. See Paper~II for the details of the simulation,
including the rise profile $h_\mathrm{a}(t)$ and the choice of the initial
equilibrium.}
\label{f:comparison2}
\end{figure}

\begin{figure}[t]                                               
\centering
\includegraphics[width=.7\textwidth]{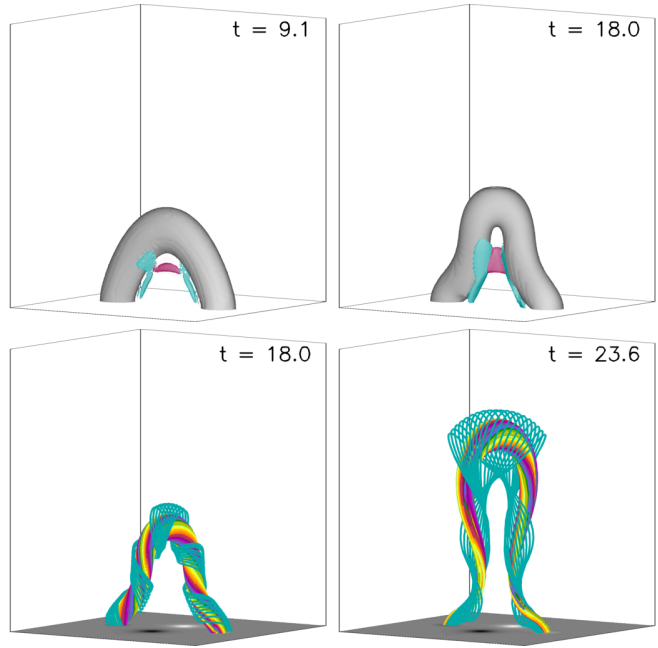} 
\caption{
Formation of the helical and vertical current sheets, and beginning
destruction of the vertical current sheet by the approaching legs of the
kinking flux rope shown in Figure~\ref{f:comparison2}. Isosurfaces of
current density $|{\bf J}({\bf x},t)|$ are displayed in the top panels;
they show the current channel in the core of the flux rope (gray), the
helical current sheet (cyan), and the vertical current sheet (red). The
isosurfaces show only the bottom part of the helical current sheet,
where the current density is highest. Field lines of the current density
${\bf J}({\bf x},t)$ are therefore plotted in the bottom panels to show
the path of the whole helical current sheet (cyan). The core of the flux
rope is displayed in these plots in the same manner as in
Figure~\ref{f:comparison2}. The final panel displays field lines in the
inflow region of the reconnection in the helical current sheet (between
the current sheet and the surface of the current channel) because part
of the field lines \emph{in} the current sheet already connect to the
ambient field at this time, as a result of the reconnection. A cubic
inner subvolume of the simulation box is displayed, and the magnetogram,
$B_z(x,y,0,t)$, is included in the field line plots.}
\label{f:hCS}
\end{figure}

\begin{figure}[t]                                               
\centering
\includegraphics[width=.7\textwidth]{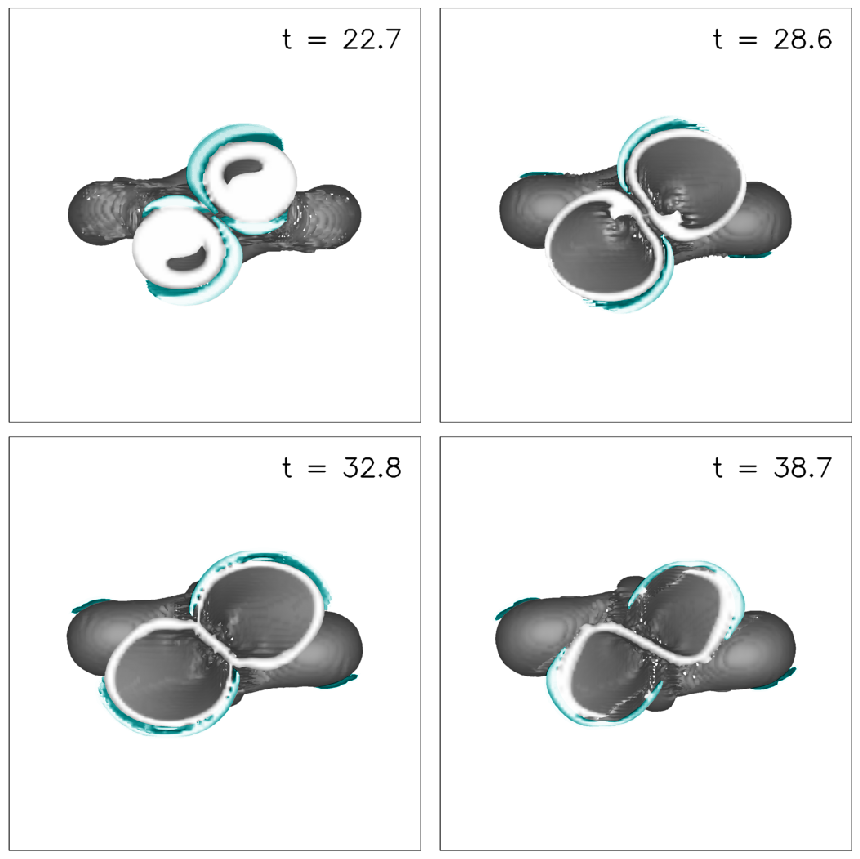}  
\caption{
Approach and reconnection of the helical current sheet (cyan) and of the
legs of the kinking flux rope (gray) shown in
Figures~\ref{f:comparison2}--\ref{f:hCS}, visualized by isosurfaces of
current density. Vertical views at the volume below a horizontal cut
plane at the height of 1.5 times the initial flux rope apex height $h_0$
are shown. The current channel of the employed Titov-D\'emoulin flux
rope has the property that the highest current density is located at its
surface. This property amplifies in the course of the dynamical
evolution, leading to an apparently hollow structure of the current
channel. However, current is flowing in the entire cross section of the
channel. See Fig.~4 in Paper~II for perspective views of a similar
simulation run.}
\label{f:reconnection}
\end{figure}

The 17~GHz emission of the brightest compact source is unpolarized and the
spectral index $\alpha$ between 17 and 34~GHz increases from zero at 02:16~UT
to $\alpha=2.1$ at 02:17~UT. This shows that the optically thin free-free
emission at 02:16~UT changes into optically thick free-free emission at
02:17~UT. In the optically thin case, we can estimate the kinetic temperature
of the electrons in the thermal background plasma from the relation $T_e=(0.2 d
n_e^2/(f^2 T_\mathrm{b}))^2$, where $d$ is the source depth, roughly estimated
from the source area $A$ as $d\sim A^{1/2}$, $f$ is the observing frequency,
and $T_\mathrm{b}$ is the observed brightness temperature (see the relations
(7), (13), and (21) in Dulk, 1985). The kinetic temperature in the source at
02:16~UT thus obtained is $T_e\approx2\times10^5$~K.

The brightness temperature and area of the final compact source plotted in
Figure~\ref{f:timeprofiles} are seen to vary, essentially in anti-phase, with a
characteristic time scale of $\sim\!40$~s. The amplitudes of the variation are
$T_\mathrm{b\,max}/T_\mathrm{b\,min}\approx1.3$ and
$A_\mathrm{max}/A_\mathrm{min}\approx2$. We check whether these values are
consistent with the assumption of perpendicular adiabatic compression and
rarefaction. Such behavior was observed in MHD simulations of plasmoids produced
by tearing and coalescence of magnetic islands in a current sheet (see, \eg,
Fig.~5 in \opencite{Schumacher&Kliem1997}). The characteristic time scale for
the oscillations of coalesced islands was found to be a few
($\sim\!3\mbox{--}5$) Alfv\'en transit times across the island in the direction
of the current sheet. From Figure~\ref{f:timeprofiles} we have a source
diameter of $\sim\!25$~Mm. Using $V_A=1000$~km\,s$^{-1}$ as a typical
value for the Alfv\'en velocity in the lower corona, an Alfv\'en transit
time of 25~s is obtained, a factor 2--3 larger than the value suggested by the
above simulation. Given that further assumptions and idealizations enter
the comparison
(\eg, the 2D idealization of the simulation), we consider this level of
mismatch to be still consistent with the assumption of perpendicular adiabatic
compression and rarefaction. This process leaves $d\approx const$ with
$n \propto A^{-1} \propto B$, where
$B$ is the field strength in the plasmoid. The first adiabatic invariant states
for the perpendicular temperature $T_\perp/B=const$. For a perpendicular
adiabatic compression which reduces $A$ by a factor 2, as estimated from
Figure~\ref{f:timeprofiles}, the density and field strength are doubled, while
the temperature $T=(2T_\perp+T_\parallel)/3$ rises by a factor 5/3, not too
different from the observed increase by a factor 1.3. Here we have assumed $T_e
\approx T_\mathrm{b}$, \ie, optically thick emission, which is appropriate for
the time range of Figure~\ref{f:timeprofiles}, as discussed above. In summary,
both the period and the amplitude variations, although not giving a clear
agreement, are consistent with the assumption of perpendicular adiabatic
compression of a plasmoid within the substantial uncertainties of the
estimates.

The dynamic radio spectrum in Figure~\ref{f:dynamic_spectrum} further shows a
group of fast-drift (type III) bursts starting in the 500--200~MHz range in the
interval 02:15--02:17~UT, which also reveal the acceleration of nonthermal
electrons at about the time the cross in the microwave source appeared at the
limb and through the subsequent phase characterized by moving compact microwave
and hard X-ray sources. Additionally, they indicate that a minor part of the
nonthermal electrons could escape along open field lines.

Figure~\ref{f:schematic} summarizes the evolutionary sequence of the main and
compact microwave sources as suggested by the observations. The evolving shape
of the source is compared in Figure~\ref{f:comparison2} with a kink-unstable
erupting flux rope, seen from a similar perspective. The overall qualitative
agreement, due to the expansion and strong writhing of the flux rope in the
course of its rise, is apparent.

This simulation, introduced in detail in Paper~II, also supports the
interpretations that the formation of compact sources, which propagated along
the legs to the top of the erupting structure, resulted from the interaction of
the legs of a kinking flux rope near the observed crossing point, and that
these sources were plasmoids. Figure~\ref{f:hCS} shows that the kinking and
rising flux rope triggers the formation of a helical current sheet in addition
to the formation of the vertical (flare) current sheet. The helical current
sheet wraps around the legs and passes over the upper section of the flux rope
in the interface to the ambient field. As the flux rope legs approach each
other, the vertical current sheet shrinks and the two layers of the helical
current sheet come into contact in its place. In the height range where this
happens, they reconnect with each other, transiently forming a single-layer
current sheet, which is subsequently squeezed and destroyed between the
partially merging flux rope legs (Figure~\ref{f:reconnection}). Thus,
``standard flare reconnection'' in the vertical current sheet is superseded in
the simulation by reconnection involving the helical current sheet and the flux
rope legs.

The two layers of the helical current sheet begin to come into contact
and to reconnect at $t\approx20$ Alfv\'en times. This leads to the
complete removal of the vertical current sheet by $t\approx24$.
Subsequently, the flux originally in the legs of the rope reconnects,
partly between the legs, partly with the ambient field (see Fig.~3 in
Paper~II for an illustration). The transition of the reconnection from
the vertical to the helical current sheet thus occurs between the second
and third snapshot pair in Figure~\ref{f:comparison2}, corresponding to
$\sim$~02:14:30--02:15:00~UT. The helical current sheet becomes the site
of dominant reconnection in the system in this period and remains so for
a period of $\lesssim\!10$ Alfv\'en times  until the current channel in
the core of the flux rope, which is seen in Figures~\ref{f:hCS} and
\ref{f:reconnection}, begins to reconnect in the full height range of
its approach. Thus, in both data and simulation, the reconnection in the
helical current sheet is dominant just after the upward acceleration of
the ejection has reached its peak, including the beginning propagation
phase after the main acceleration (see the rise profile of the flux rope
in Paper~II). The corresponding period in the observations extends from
the time the hard X-ray source began its rapid ascend
($\approx$\,02:14:45~UT) till and somewhat beyond the formation of the
strongest compact microwave sources ($\approx$\,02:15:30~UT). From this
timing and from the path of the hard X-ray and compact microwave
sources---from the crossing point, along the legs, and to the apex of
the flux rope---we conjecture that reconnection in the helical current
sheet formed plasmoids which subsequently propagated along the helical
current sheet and became visible as these radiation sources by trapping
nonthermal particles.

This does not exclude that the major part of the emitting nonthermal
particles was actually accelerated by reconnection in the vertical
current sheet, as suggested by the standard flare picture. The major
part of the hard X-rays and microwaves is emitted simultaneously with
the main CME acceleration (compare Figure~\ref{f:lightcurves} and Fig.~1
in \opencite{Hudson&al2001} with Figure~\ref{f:accel}). The main
acceleration of fast CMEs is often synchronized with flare reconnection,
thought to occur in the vertical current sheet (\eg,
\opencite{Lin&Forbes2000}; \opencite{Qiu&al2004};
\opencite{Zhang&Dere2006}), and this is the case also in the simulation
shown in Figures~\ref{f:comparison2}--\ref{f:reconnection}. Most of the
flux rope's acceleration is complete before the helical current sheet
begins to reconnect strongly. However, the comparison of the data with
the simulation suggests that it is the replacement of the vertical
current sheet by the helical current sheet which allows a large part of
these particles to become trapped and propagate to the top of the
erupting flux rope. The acceleration of particles continues past the
suggested transition of reconnection from the vertical to the helical
current sheet, as evidenced by continuing strong microwave emission from
the limb source (near the crossing point) and by the creation of
fast-drift bursts through 02:17~UT.

Finally, the reconnection between the legs of the kinking flux rope commences
and remains at low heights (see Figures~\ref{f:comparison2}--\ref{f:hCS} and
the detailed consideration in Paper~II). This is consistent with the location
of peak emission in the main microwave source near the limb throughout the
event.

\section{Discussion}
\label{s:discussion}

The inverse gamma shape of the main microwave source, revealed by the 17~GHz
images in Figure~\ref{f:leg_crossing}, suggests that the erupting flux was loop
shaped and experienced a strong writhing in the course of its rise. This
writhing is consistent with both possible causes, the occurrence of the helical
kink instability, and the lateral forces acting on the legs of the loop due to
a shear field component of the ambient field if the loop is accelerated by
another process.

However, the strongest microwave and hard X-ray sources provide evidence that
the legs of the rising flux loop did interact near the point of their crossing
(as seen in the projection onto the plane of the sky). Magnetic reconnection is
the prime candidate for such interaction, which is confirmed by the numerical
simulations in Figures~\ref{f:comparison2}--\ref{f:reconnection} and Paper~II.
These simulations demonstrate that the legs of a writhing flux loop approach
each other and reconnect only if the deformation is a double helix (normalized
axial wavenumber $k'\sim2$). This cannot be achieved through the writhing by a
shear field component and implies the occurrence of the helical kink
instability of a flux rope that has an initial twist of $\approx\!6\pi$ or
higher.

The observation data do not allow to determine whether such a twist
could accumulate prior to the eruption, or was built up in the early
phase of the eruption.

Three pieces of evidence suggest strongly that nonthermal particles were
accelerated, and compact propagating sources were formed, at or near the
crossing point. First, the point of peak brightness of the microwave source
stayed close to the crossing point throughout the event. Second, part of the
nonthermal particle population became trapped in compact microwave and hard
X-ray sources whose first appearance and motion suggest an origin associated
with the legs of the erupting flux loop, near the crossing point, and near or
shortly after the time of peak energy release rate (peak hard X-ray flux). A
coincidental association with the rise of the crossing point above the limb and
the hard X-ray peak, implying an origin  of the compact sources at earlier
times from below the crossing point, cannot be excluded, due to the occultation
of the lower parts of the flaring volume. However, we regard such a coincidence
to be far less likely. The simultaneous occurrence of the first two compact
microwave sources in the two legs slightly above the crossing point, with
similar distances and velocities, argues against a coincidence in particular.
Third, the escaping part of the nonthermal particles was first detected through
its fast-drift emission (type III bursts) at meter wavelengths at the time the
crossing point had appeared at the limb.

The subsequent rapid fading of the upper part of the inverse gamma
shaped main microwave source is consistent with the scenario that the
interaction of the flux rope legs caused a breakup of the original rope,
as suggested for another event by \inlinecite{Cho&al2009} and as also
indicated by the numerical simulations in Paper~II. However, the
fading can also be explained simply as a  consequence of the upward
expansion of the source, which causes the field strength and density in
the flux loop to decrease. The occultation of the source's lower
part prevents a decision between these possibilities.

The interpretation of the moving compact microwave and hard X-ray sources as
plasmoids is not proven by the available data, but it is supported by several
aspects. First, the localization in the course of propagation across the
substantial height range of several 100~Mm requires a high efficiency of
particle trapping, which is a known property of plasmoids. Alternative
structures---solitons or shocks---are likely less efficient in trapping the
nonthermal particles. Second, the observed parameters of the source
oscillations are roughly consistent with the assumption of perpendicular
adiabatic compression and rare\-faction. This is a typical behavior of
plasmoids created by island coalescence in a current sheet. Third, the region
of plasmoid formation and their trajectories correspond to a current sheet
computed in the model: the helical current sheet. It has been demonstrated in a
long series of investigations that plasmoids are formed and propagate in
reconnecting current sheets, due to the tearing and coalescence instabilities
(see, \eg, \opencite{Leboeuf&al1982}; \opencite{Matthaeus&Lamkin1986};
\opencite{Scholer&Roth1987}; \opencite{Schumacher&Kliem1996};
\opencite{Magara&al1997}; \opencite{Shibata&Tanuma2001};
\opencite{Loureiro&al2007}; \opencite{Karlicky&Barta2007};
\opencite{Barta&al2008}; \opencite{Bhattacharjee&al2009};
\opencite{Barta&al2010}; \opencite{Edmondson&al2010}). All these investigations
were performed for the geometry of a plane (single or double) current sheet,
but they span a very wide range of parameters and applications and use a
variety of different methods; the most recent one is even three-dimensional.
Therefore, the formation and acceleration of plasmoids is established as an
important process in the dynamics of reconnecting current sheets. The
simulations presented in Figures~\ref{f:comparison2}--\ref{f:reconnection} and
Paper~II are fully three-dimensional and model the instability of a flux rope
at active-region scales, so that they necessarily lack the resolution required
to demonstrate the conjectured process of plasmoid formation in the thin
helical current sheet. However, the general evidence provided by the quoted
investigations (and references therein) supports our conjecture (see also
Paper~II).

Furthermore, as shown by \inlinecite{Barta&al2008}, plasmoids formed in
the reconnection region of an eruptive flare move upward if the magnetic
field surrounding them diverges in the upward direction. This is the
natural situation in the solar atmosphere and explains why most
plasmoids move upward. For the same reason, a plasmoid that has moved
to the top of an erupting flux rope (in the helical current sheet in the
present case) will remain there and continue its upward motion joint
with the rope.

Can plasmoid formation in the vertical current sheet according to the
standard scenario \cite{Ohyama&Shibata1998} also be consistent with the
observations of the event? In this case it remains unclear how the
plasmoids could find a path to the top of the rope which first follows
the legs. The field lines reconnected in the vertical current sheet pass
near the apex of the rope only if the shear field component of the
ambient field is weak, but in this case the plasmoids propagate
essentially vertically between the legs without approaching them. If the
shear field component is significant, then the field lines reconnected
in the vertical current sheet do wrap around the legs of the flux rope
(opening a path for plasmoids and accelerated particles to the legs),
but from there they continue downward to the photosphere, \emph{not} to
the apex of the flux rope (see, \eg, Figs.~3 and 4 in
\opencite{Kliem&al2004} for a field line plot of such a configuration).

\begin{figure}[t]                                               
\centering
\includegraphics[width=.7\textwidth]{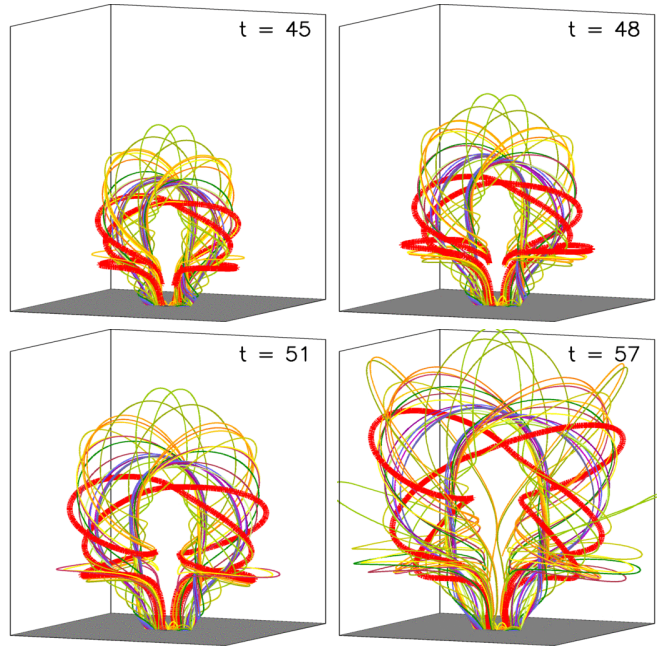} 
\caption{
Formation of a relatively localized sharp bend in two symmetrical field
lines at the surface of the flux rope legs by reconnection with the
ambient field (thick red field lines). The bends move upward along the
legs as the flux rope rises and the newly reconnected field lines
relax.}
\label{f:alternative}
\end{figure}

Finally, we consider an alternative mechanism for the formation of
localized, traveling structures in the legs of an erupting flux rope. If
the reconnection in the vertical current sheet under the rising flux
rope does not only merge pairs of field lines that are both external to
the rope, but also field lines in the legs of the rope with field lines
of the ambient field, then relatively sharp bends can temporarily occur
in the reconnected field lines in the legs. The bends move upward along
the legs as the newly reconnected field lines relax to a less bent
shape. Special requirements on the driving mechanism of the eruption,
for example a high initial twist, are not implied.
Figure~\ref{f:alternative} gives an illustration taken from a run that
does not involve the helical kink instability (from
\opencite{Schrijver&al2008a}); note that the bends appear sharper than
in reality due to the projection. While this mechanism is consistent
with the formation site and path of the compact microwave and hard X-ray
sources in the analyzed event, it does not explain the efficient
trapping of the particles. The field line plots in
Figure~\ref{f:alternative} show that the bends quickly lose their
localization in the course of the upward propagation, so that any
accelerated particles can easily populate large parts of these field
lines. Indications for the existence of localized structures and
oscillations at the apex of the flux rope are completely lacking.
Therefore, the observations of the event do not support such a scenario.

Thus, we suggest that a scenario for the occurrence of plasmoids
may be realized in some flares, which differs from the standard scenario
of formation and propagation in the flare current sheet as suggested by
\inlinecite{Ohyama&Shibata1998}. According to the new scenario,
plasmoids can also form and propagate in a helical current sheet wrapped
around strongly distorted, interacting flux ropes. In the case studied
here, which requires a rather high initial twist in the flux rope,
the rope interacts with itself. Plasmoids may be excited in a
similar manner if a highly distorted flux rope interacts with ambient
flux, which would not require such high twist.

\section{Conclusions}
\label{s:conclusions}

The microwave data of the eruptive flare on 18 April 2001 in AR~9415, which
originated $\approx\!26^\circ$ behind the limb and was associated with a fast
CME, suggest that the event involved the helical kink instability of a highly
twisted flux rope. This is based on the inverse gamma shape of the microwave
source as a whole and on evidence that the legs of the rising flux loop
interacted (reconnected) at or near their crossing point seen projected in the
plane of the sky. No other process is known that could produce both phenomena
simultaneously. The evidence for leg-leg interaction derives from the
acceleration of energetic, nonthermal particles and from the formation of
compact sources at or near the crossing point, which subsequently move upward
along the legs of the rising flux loop.

The efficient trapping of fast particles in the compact moving microwave
and hard X-ray sources suggests that these sources are plasmoids
(magnetic islands in a current sheet) by nature. This interpretation is
supported by their oscillations and is consistent with the motion away
from their origin. Their path along the upper part of the inverse gamma
source and the coalescence of multiple blobs into a final compact source
at the top of the inverse gamma source are consistent with formation
and propagation in the helical current sheet formed by the kink
instability. Thus, the event suggests that plasmoids may form in current
sheets steepened by the lateral displacement of flux tubes, in addition
to the standard scenario of formation in the flare current sheet below
erupting flux.

\acknowledgements The authors thank the referee for constructive comments which
helped to improve the paper and T. T{\"o}r{\"o}k for a careful reading of
the manuscript. M. Karlick\'y should like to acknowledge the support from the
Nobeyama National Observatory and the kind hospitality of
Prof.~K.~Shiba\-sa\-ki and his staff during his stay at Nobeyama. Furthermore,
he thanks Dr.~M.~Shimojo for the maps presented in
Figures~\ref{f:plasmoid_merging} and \ref{f:comparison2}. We acknowledge the
HiRAS radio spectrum from the Hiraiso Solar Observatory, the use of the LASCO
CME catalog, generated and maintained at the CDAW Data Center by NASA and The
Catholic University of America in cooperation with the NRL, and the
\textsl{SOHO} archive of EIT data.
\textsl{SOHO} is a project of international cooperation between ESA and NASA.
This study was supported by
Grant 300030701 of Grant Agency of the Czech Academy of Sciences,
by the DFG, an STFC Rolling Grant,
and NASA grants NNH06AD58I and NNX08AG44G.

\end{article}
\end{document}